\let\orilabel\label
\documentclass[aps,twocolumn]{revtex4-2}
\let\label\orilabel

\usepackage{amsfonts,amsmath,bm,amssymb,revsymb,color,braket}
\usepackage{xcolor}
\usepackage{graphicx}
\usepackage[makeroom]{cancel}
\definecolor{bostonuniversityred}{rgb}{0.8, 0.0, 0.0}
\definecolor{dukeblue}{rgb}{0.0, 0.0, 0.61}
\definecolor{ao(english)}{rgb}{0.0, 0.5, 0.0}
\definecolor{darkmagenta}{rgb}{0.55, 0.0, 0.55}
\definecolor{armygreen}{rgb}{0.29, 0.33, 0.13}
\definecolor{coquelicot}{rgb}{1.0, 0.22, 0.0}
\definecolor{fucsiak}{rgb}{0.4, 0.08, 0.4}
\definecolor{airforceblue}{rgb}{0.36, 0.54, 0.66}
\definecolor{applegreen}{rgb}{0.55, 0.71, 0.0}
\definecolor{awesome}{rgb}{1.0, 0.13, 0.32}
\definecolor{burgundy}{rgb}{0.5, 0.0, 0.13}
\definecolor{cobalt}{rgb}{0.0, 0.28, 0.67}
\definecolor{aqua}{rgb}{0.0, 1.0, 1.0}
\definecolor{blue-green}{rgb}{0.0, 0.87, 0.87}
\usepackage[unicode=true,bookmarksnumbered,colorlinks=true, citecolor=burgundy, linkcolor=blue, urlcolor=cobalt]{hyperref}

\begin{document}

\title{Maintenance of a columnar vortex by inertial waves in rotational turbulence}

\author{Nikolay A. Ivchenko}

\affiliation{Landau Institute for Theoretical Physics,
Russian Academy of Sciences,\\1-A Akademika Semenova av.,
142432 Chernogolovka, Russia}

\affiliation{National Research University Higher School of
Economics, Laboratory for Condensed Matter Physics, 101000 Moscow, Russia}

\email{ivchenko@itp.ac.ru}

\author{Sergey S. Vergeles}

\affiliation{Landau Institute for Theoretical Physics,
Russian Academy of Sciences,\\1-A Akademika Semenova av.,
142432 Chernogolovka, Russia}

\affiliation{National Research University Higher School of
Economics, Faculty of Physics, Myasnitskaya 20, 101000
Moscow, Russia}

\email{ssver@itp.ac.ru}

\author{Daniil D. Tumachev}

\affiliation{Landau Institute for Theoretical Physics,
Russian Academy of Sciences,\\1-A Akademika Semenova av.,
142432 Chernogolovka, Russia}

\author{Sergey V. Filatov}

\affiliation{Landau Institute for Theoretical Physics,
Russian Academy of Sciences,\\1-A Akademika Semenova av.,
142432 Chernogolovka, Russia}
\affiliation{Osipyan Institute of Solid State Physics, Russian Academy of Sciences,  \\ Chernogolovka, Moscow region, 142432 Russia}
\begin{abstract}
    We develop a theory that determines the radial profile of the mean velocity in a coherent geostrophic vortex forming in a turbulent rapidly rotating fluid. Following the conditions of our experiments, we assume that the flow in the vortex is sustained by the absorption of short-wavelength inertial waves arriving from the vortex periphery. The nonlinear interaction between waves is assumed negligible in the theory. Comparison with experimental data supports the validity of the developed model.
\end{abstract}

\maketitle

\section{Introduction}

The flow of a  rapidly rotating fluid can be decomposed into two components: a quasi-two-dimensional geostrophic flow and inertial waves~\cite{davidson2024dynamics}. As the flow becomes turbulent, an inverse energy cascade establishes in geostrophic flow, supplemented by energy transfer from inertial waves. Thus, long-living geostrophic vortices can form within the system, which are statistically stable structures having flow sustained by an inflow of energy from smaller-scale turbulent pulsations~\cite{godeferd2015structure}. The interest lies not only in the statistical analysis of the characteristics of the entire flow as a whole, but also in determining the mean flow's structure in such coherent vortices.

The pioneering observation of a quasi-two-dimensional vortex formed in turbulent rotating system was~\cite{mcewan1976angular}, where the flow was driven by injecting water through a perforated bottom of the vessel. The first systematic study was the experiment~\cite{hopfinger1982turbulence}, with rapidly oscillating grid at the bottom of the tank as the excitation mechanism. A theoretical study~\cite{kolokolov2020structure} predicted a linear-logarithmic velocity profile for the vortex under conditions that turbulent fluctuations are forced homogeneously over the volume. Such conditions were realized in experiments~\cite{xia2009spectrally,orlov2018large} on two-dimensional flows, where this kind of coherent vortices were formed. Also, the linear-logarithmic velocity profile of the coherent vortex has been observed in numerical simulation in Ref.~\cite{doludenko2021coherent} of two-dimensional turbulent dynamics without bottom friction effects~\cite{parfenyev2021influence}. 

The vortices arising in Ref.~\cite{hopfinger1982turbulence}, were approximately aligned with the rotation axis of the system, and their lifetime was is ten to twenty own rotation periods; see Section~6 of the review~\cite{hopfinger1993vortices}. The latter is a characteristic drawback of experiments in which the flow is driven by constructions at the bottom (grid or tubes)~\cite{mcewan1976angular,ruppert2004extraction}. These structural irregularities of the horizontal boundary increases sufficiently the decay of the geostrophic flow as compared with the Ekman rate~\cite{pedlosky1996ocean, parfenyev2021influence}, that was confirmed e.g. in Ref.~\cite{morize2005decaying}. 

In the present work we develop a theory that describes the structure of the long-living vortices observed in our experiments~\cite{tumachev2023two,tumachev2024observation}, where the excitation takes place at the vertical boundaries of the vessel and the horizontal boundaries were smooth. The vortex structure is determined by the Reynolds equation for the mean flow in the vortex. The Reynolds stress appearing in this equation is formed by the statistics of the inertial waves. In general, the waves dynamics is affected by both the wave-vortex interaction and the nonlinear wave-wave interaction, though the limit of strong rotation realized in the experiments allows to neglect the latter one. The radial profiles of mean velocity $U(r)$ we obtained are a result of the radially distributed absorption of specific part of converging cylindrical waves and of the reflection from the vortex flow of the remaining part of waves. The profile depends on the statistical properties of the wave ensemble and the fluid rotation speed. It is quite nontrivial, so its approximation by e.g. Rankine vortex model~\cite{giaiotti2006rankine} would lead to an oversimplification when describing the experimental observations~\cite{tumachev2023two,tumachev2024observation}.

\section{Quasi-linear approximation for the dynamics} 
\label{sec:gen}

We consider the flow of an incompressible fluid in a box of height $H$ that is rotating with an angular velocity $\Omega$ assumed to be directed along the vertical axis $Oz$. In the total velocity field of the fluid $\bm v$, we single out a geostrophic component $U(t;x,y)$, which depends on the horizontal coordinates only and is directed perpendicular to the rotation axis, and a part $\bm u$ with fast dynamics in time, which depends essentially on the vertical coordinate: $\bm v =\bm U+\bm u$. We assume the rotation to be fast; the field $\bm u$ then describes an ensemble of inertial waves. When averaging over the fast wave oscillations, field $\langle \bm u\rangle$ becomes equal to zero.

In the model, we presume the geostrophic flow inside the vortex to be axisymmetric following the theory~\cite{kolokolov2020structure}. Thus, its velocity has the form $\bm U= U(r)\bm e_\varphi$ in cylindrical coordinate system $(r,\varphi,z)$ ($z$-axis coincides with the vortex axis), where $\bm e_\varphi$ is the unit vector in the azimuthal direction. The steady flow of the vortex is governed by the Reynolds equation, which is the Navier-Stokes equation averaged over the fast turbulent fluctuations~\cite{parfenyev2021influence}:
\begin{equation}
    \alpha U 
    +
    \frac{1}{r^2} \partial_r 
    \Big(r^2\Pi^{r\varphi}-\nu r^2\Sigma\Big)
    =
    0,
    \ \ 
    \Pi^{r\varphi}=\left\langle u^r u^\varphi\right\rangle,
\label{gen:mean_flow}
\end{equation}
where the angular brackets $\langle\ldots\rangle$ denote averaging over the ensemble of inertial waves, $\nu$ is the kinematic viscosity, and $\alpha = 2\sqrt{\nu\Omega}/H$ is the bottom friction coefficient~\cite{pedlosky1996ocean,parfenyev2021influence}. The local shear rate $\Sigma=U^\prime - U/r$ (here and below a prime denotes the derivative with respect to~$r$) characterizes the differential rotation of the fluid in the vortex. Below, it will be more convenient to describe the flow in terms of its angular velocity $W=U/r$, so that $\Sigma=r W^\prime$.

The waves are presumed to be excited at the periphery and then propagate into the body of the vortex. At the periphery, where $U$ tends to be zero, the waves in the ensemble possess spatially homogeneous and isotropic statistics with a characteristic wavevector magnitude $k_f$. Let us choose a certain distance $R_0$ from the vortex axis that conventionally separates body of the vortex from the periphery. At $r>R_0$ the mean kinetic energy of waves in the ensemble $\langle \bm u^2\rangle /2$ is proportional to the energy flux $f$ per unit mass (of dimension $\,\mathrm{cm}^3/\,\mathrm{s}^3$) that waves carry through the cylindrical surface $r=R_0$. This flux sets the maximum power the vortex can possibly extract from turbulent pulsations in order to sustain its own flow. Earlier theoretical works~\cite{kolokolov2020structure,parfenyev2021influence} employed different parameter of the energy transfer from waves to the mean flow, which is the power per unit mass $\epsilon$ (of dimension $\,\mathrm{cm}^2/\,\mathrm{s}^3$). If one assumes, as a matter of normalization, that the energy transfer is uniformly distributed throughout the volume $r<R_0$, the relation between these two quantities is
\begin{equation}
	2\pi R_0 f =\pi R_0^2 \epsilon.
\label{gen:epsilon}
\end{equation}

In order for the turbulent pulsations to be considered as inertial waves, their nonlinear interaction should be weak compared to the wave Coriolis force, that is, the Rossby number $\mathrm{Ro}\sim\left|\nabla \bm u\right|/2\Omega$ should be small. In the weak-turbulence regime the mean kinetic energy of the waves is proportional to $\langle \bm u^2\rangle \sim \sqrt{\epsilon\Omega}/k_f$~\cite{galtier2003weak}, so the requirement for the wave Rossby be small number leads to the inequality
\begin{equation}
	k_f \sqrt{\epsilon/\Omega} 
    \ll 
    \Omega.
\label{gen:inequality1}
\end{equation}
The left‑hand side of~(\ref{gen:inequality1}) is the rate~$1/\tau_{\mathrm{tr}}$ of nonlinear wave-wave interaction~\cite{galtier2003weak}. We assume this effect to be negligible compared to the nonlinear interaction of waves with the geostrophic flow. The dynamics of the field $\bm u$ is then governed by the linearized Navier–Stokes equation~\cite{ivchenko2025absorption}.  The requirement that~$1/\tau_{\mathrm{tr}}$ is small compared both with the rate $\Sigma$ which characterizes the interaction of the waves with the geostrophic vortex and with the inverse time $\Omega/k_f R_0$ it takes a wave to travel through the body of vortex leads to the inequalities
\begin{equation}
    k_f \sqrt{\epsilon/\Omega} 
    \ll 
    \Sigma, \ \Omega /k_f R_0.
\label{gen:inequality2}
\end{equation}
The size of the vortex is assumed to be large compared with the wavelength~$1/k_f$, and the Rossby number of the vortex $\Sigma/2\Omega\ll 1$, so each of the two inequalities in (\ref{gen:inequality2}) is stronger than (\ref{gen:inequality1}).

\section{Inertial waves in the vortex's flow} 
\label{sec:cyl}

\begin{figure}[t]

{\includegraphics[width=\linewidth]{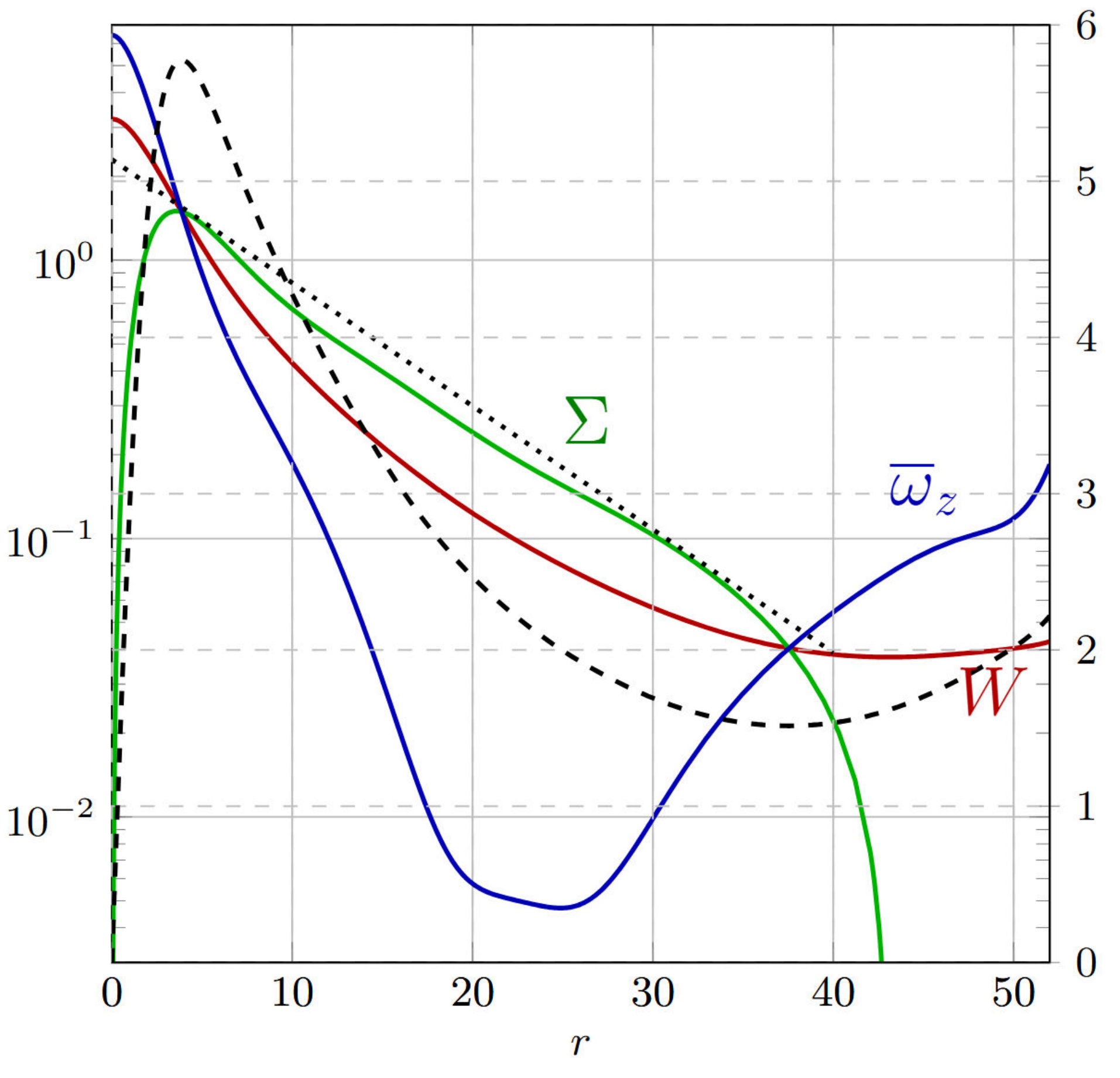}}

\caption{Mean flow profile of the cyclone for experiment at~$\Omega=2.63\,\text{s}^{-1}$, which is described in Section~\ref{sec:mean}. The radial dependence is plotted for azimuthal velocity~$U(r)$ measured in $\text{cm}/\text{s}$ (dashed line, right y-axis), as well as for angular rotation frequency $W(r)$, shear rate $\Sigma(r)$ and local vorticity $\overline \omega_z = r^{-1}\partial_r(rU)$ measured in $1/\text{s}$ (logarithmic scale, left y-axis). The dotted line is the exponential dependence $\propto \exp(-0.1 r)$, that corresponds to the intermediate asymptotic of $\Sigma(r)$.} 

\label{fig:profile}
\end{figure}

The equations for wave propagation on the background
of an axisymmetric steady vortex are homogeneous in the variables $t,\varphi$ and $z$. Therefore, we pass to the frequency~$\omega$, the azimuthal number~$m$, and the vertical wavenumber~$k_z$. The Equation (A6) of Ref.~\cite{ivchenko2025absorption} for the reduced pressure $p$ of the wave is 
\begin{equation}
    \begin{gathered}
    \widetilde{\omega}^{2}\Delta p+2\widetilde{\Omega}\left(2\widetilde{\Omega}+\Sigma\right)k_{z}^{2}\,p=0,
    \\
    \Delta=\partial_r^2+\frac{1}{r}\partial_r -k_{{\scriptscriptstyle \perp}}^{2}(r),
    \end{gathered}
\label{cyl:p_shortwave}
\end{equation}
in the inviscous limit and in the short-wave approximation, where $\widetilde\omega(r)=\omega-mW$ the Doppler‑shifted relative frequency of the wave, $k_{{\scriptscriptstyle \perp}}^{2}(r)=k_z^2+m^2/r^2$ is the squared wavenumber in the plane transverse to the radial direction and $\widetilde\Omega(r)=\Omega+W$ is the total local angular velocity of fluid rotation. The wave equation~(\ref{cyl:p_shortwave}) gives the dispersion relation
\begin{equation}
    \widetilde\omega
    =   \sqrt{2\widetilde\Omega\big(2\widetilde\Omega+\Sigma\big)}\,
    s k_z\Big/
    \sqrt{k_r^2 + k_{{\scriptscriptstyle \perp}}^{2}}\,,
\label{cyl:WKB-disp}
\end{equation}
in which $s=\pm 1$ stands for polarization of the wave and $k_r(r)$ is the wavenumber in the radial direction. Equation~(\ref{cyl:WKB-disp}) may be regarded as an equation for $k_r$. The  velocity components of a cylindrical harmonic in horizontal plane, $u^r\equiv v$, $u^\varphi\equiv u$, are expressed in terms of the pressure via
\begin{gather}
    v
    =
    \frac{i\left(\widetilde{\omega}p^\prime-2\widetilde{\Omega}pm/r\right)}
    {4\widetilde{\Omega}^{2}+2\widetilde{\Omega}\Sigma-\widetilde\omega^2},
    \ 
    u
    =
    \frac{\left(2\widetilde{\Omega}+\Sigma\right)p^{\prime}-\widetilde{\omega}pm/r}
    {4\widetilde{\Omega}^{2}+2\widetilde{\Omega}\Sigma-\widetilde{\omega}^{2}}.
    \label{cyl:uv-via-p}
\end{gather}
The applicability of the short-wave approximation requires the following inequalities to be hold:
\begin{equation}
	k_r^\prime\ll k_r^2;
    \qquad 
    \left(\ln W\right)^\prime,\:\left(\ln \widetilde\Omega\right)^\prime\ll k_r.
\label{cyl:shortwave_cond}
\end{equation}

Let us examine the behavior of the solution $k_r(r)$ of equation~(\ref{cyl:WKB-disp}) for a converging cylindrical wave. A typical cyclone vortex's flow profile based on the experimental data~\cite{tumachev2023two,tumachev2024observation} is presented in Fig.~\ref{fig:profile}. According to the data, we assume that the angular velocity $ W(r)$ is a monotonically decreasing function in its absolute value. At large distances, where the mean flow is absent ($W\rightarrow 0$), the total wavenumber $\sqrt{k_r^2+k_{{\scriptscriptstyle \perp}}^{2}}$ in the dispersion relation~(\ref{cyl:WKB-disp}) reaches a fixed value, which we denote as $k$. The absolute value of the projection of the wavevector onto the horizontal plane equals to $q$, where $q^2=k^2-k_z^2=k_r^2+m^2/r^2$. Hence, at the largest distances $r\gg |m|/q$, the radial wavenumber is $k_r=q$. As $r$ decreases, one of two features of the solution $k_r(r)$ is encountered first. If it is the turning point $r=r_t$ where $k_r(r_t)=0$, the converging cylindrical wave undergoes reflection there, after that is turns into a diverging wave. The other feature is the critical layer at~$r=r_\ast$ where $\widetilde\omega(r_\ast)=0$, which means the singularity for a wavenumber $k_r\rightarrow \infty$ according to~(\ref{cyl:WKB-disp}). The wave undergoes resonant absorption by the mean flow in the critical layer~\cite{booker1967critical,fabrikant1998propagation,dileoni2015}. The efficiency of this absorption grows as the local Rossby number of the wave $\rho$ decreases; see Eq.~(36) of Ref.~\cite{{ivchenko2025absorption}}. Due to the assumed smallness of the Rossby number of the vortex, $\Sigma/2\Omega\ll1$, $\rho$ is expected to be small for a typical wave. The condition $\rho\ll 1$ means that the first inequality in~(\ref{cyl:shortwave_cond}) holds, and the transmitted and reflected waves can be neglected as their amplitudes are small in the parameter~$e^{-\pi /\rho}\ll 1$. With the dispersion relation~(\ref{cyl:WKB-disp}) this condition can be written in the form
\begin{equation}\label{cyl:CL_invisc}
    \rho
    =
    \frac{2\Omega}{\sqrt{2\widetilde\Omega(2\widetilde\Omega+rW^\prime)}}
    \cdot
    \frac{\left|mW^\prime\right|}{k |\omega|}
    \ll 1.
\end{equation}
When the Rossby number of the geostrophic flow is small, $W\ll 2\Omega$, the first factor in~(\ref{cyl:CL_invisc}) is close to unity and this condition reduces to $|m W^\prime/k \omega |\ll1$.

For a monochromatic cylindrical wave in the inviscid limit, the exact wave equations imply that the flux of angular momentum through a surface of radius $r$ associated with the wave is conserved~\cite{ivchenko2025absorption}: 
\begin{equation}
	2\pi r^2\Pi^{r\varphi}
    =
    2\pi r^2\langle {u^r u^\varphi}\rangle 
    =
    \pi r^2\mathrm{Re}\big(uv^\ast\big)
    =
    \mathrm{const}.
\label{cyl:const_flux}
\end{equation}
where the averaging is performed over the wave oscillations. The quantity~(\ref{cyl:const_flux}) is one that enters the Reynolds equation~(\ref{gen:mean_flow}). The constancy of the flux is violated near the critical layer because of wave absorption. This was first studied in Ref.~\cite{booker1967critical} for the problem of internal waves in a stratified fluid.

In Eq.~(\ref{cyl:p_shortwave}) the influence of viscosity was not taken into account. This is valid if the dissipation rate $\nu\left(k_r^2+k_{{\scriptscriptstyle \perp}}^2\right)$ is much smaller than both the wave frequency $\widetilde{\omega}$ and the rate of its interaction $\Sigma$ with the geostrophic flow in the vortex body. The first condition is violated in the vicinity of the critical layer. The absorption of an inertial wave near the critical layer in the presence of viscous effects was analyzed in~\cite{{ivchenko2025absorption}}. There was obtained a solution in the neighbourhood of the layer which is characterized by the viscous spatial scale $\eta_\mathrm{v}=\left|\nu/mW^\prime\right|^{1/3}$ (see Eqs.~(49) and (C2)). For small $\rho$, the wave decays at distances $\rho^{-2/3}\eta_\mathrm{v}$ around it. We consider the small‑viscosity approximation throughout the work, presuming that $\eta_\mathrm{v}$ is significantly smaller than other scales in the problem, and regard absorption in the critical layer as the principal mechanism for the energy and angular momentum transfer to the mean flow. In particular, we neglect viscous dissipation during wave propagation outside the critical layer.

\section{Outer ensemble of waves}
\label{sec:out}

In this Section, we establish statistical properties of the ensemble of converging cylindrical waves. At the periphery of the flow, the field $\bm u$ corresponds to the ensemble of inertial waves in a uniformly rotating fluid. We assume this ensemble to be statistically homogeneous in space and isotropic. In this case one can write the plane-wave expansion for the wave velocity field $\bm u$ and describe it in the orthonormal basis of transverse helical modes, see Refs.~\cite{sagaut2008homogeneous, kolokolov2020structure}, characterized by the wavevector $\bm k$ and helical polarization $s=\pm 1$:
\begin{equation}
	  \bm u (t, {\bm r}) 
    =
    \sum_{s=\pm 1} \int \frac{d^3 \bm k}{(2\pi)^3}
	  a_{\bm k}^s \bm h_{\bm k}^s 
   \exp\big(-i\omega^s_{\bm k}t + i\bm k\cdot {\bm r}\big),
\label{out:helical_vel}
\end{equation}
where the plane wave's frequency is 
\[
    \omega_{{\bm k}}^s =\frac{2\Omega k_z s}{k}.
\]
and $\bm h^s_{\bm k}$ are complex vectors defined in the plane orthogonal to $\bm k$ as:
\begin{equation}
    \begin{aligned}
    h_{\bm{k}}^{s}=\frac{\bm e^{(2)}-is \bm e^{(1)}}{\sqrt{2}},
    \\
    \bm e^{(1)} =\frac{\left[\bm{k}\times\bm{e}^{z}\right]}{\left|\bm{k}\times\bm{e}^{z}\right|},
    \qquad 
    \bm e^{(2)}=\frac{\left[\bm{k}\times\bm e^{(1)}\right]}{k }.
    \end{aligned}
\label{out:circular_comp}
\end{equation}

The corresponding to~(\ref{out:helical_vel}) expansion of wave pressure field takes the form:
\begin{equation}
	  p 
    = 
    \sum_{s=\pm 1} s 
    \int \frac{d^3 \bm k}{(2\pi)^3}
    \frac{ -\sqrt{2}\Omega q}{k^2} \,
    a_{\bm k}^s e^{-i\omega_{{\bm k}}^s t+i\bm k\cdot \bm r},
    \label{out:helical_press}
\end{equation}
where notation of $q^2 = k_x^2 + k_y^2$ is the same one that was introduced below the Eq.~(\ref{cyl:shortwave_cond}) for cylindrical wave.

The ensemble in the model is defined in terms of the amplitudes $a^s_{\bm k}$ of plane waves with stationary statistics. It is assumed to be homogeneous and isotropic in space and Gaussian with zero mean. The pair correlation function is
\begin{equation}\label{out:ensemble}
    \begin{aligned}
	\left\langle a^s_{\bm k} \, a^{s^\prime\, \ast}_{\bm k^\prime}\right\rangle 
    =
    \left(2\pi\right)^3\delta^{s s^\prime}\delta^{(3)}\left(\bm k - \bm k^\prime\right) 
    \cdot 
    \frac{3\pi f}{\Omega} k \Xi(k),
	\\
	\int \frac{d^3 \bm k}{\left(2\pi \right)^3}\Xi(k)=1.
    \end{aligned}
\end{equation}
The isotropy means that the normalized correlation function $\Xi(k)$ depends on the absolute value of wavevector $k$ only.

In absence of a geostrophic flow, Eq.~(\ref{cyl:p_shortwave}) for pressure of the cylindrical wave becomes exact. Its solution, which remains finite at the origin, is the Bessel function $J_m(qr)$ of order $m$. A single harmonic is a cylindrical standing wave defined by a set of numbers $\bm \chi=\left(q,m,k_z,s\right)$. The velocity components in explicit form are~\cite{zhang2017theory}:
\begin{gather}
	v= b_{\bm\chi}\left[\left(1-s\cos\theta\right)J_{m-1}+\left(1+s\cos\theta\right)J_{m+1}\right](qr),
	\nonumber
	\\
	u=i b_{\bm\chi}\left[\left(1-s\cos\theta\right)J_{m-1}-\left(1+s\cos\theta\right)J_{m+1}\right](qr),
	\nonumber
	\\
	w=2is\sin\theta b_{\bm\chi} J_{m}(qr),\quad \cos\theta=\frac{k_z}{k}.
	\label{out:cylinder_basis}
\end{gather}
The corresponding expansion of the pressure in this basis is: 
\begin{equation}
    \begin{aligned}
	p(t,r,\varphi,z)
    =
    \sum\limits_{s=\pm 1} 
    \int \frac{d k_z}{2\pi}
    \int \limits_0^\infty q \,dq \sum_m 
    \frac{4i\Omega q}{ k^2} 
    \\
    e^{im\varphi + ik_z z - i\omega_{{\bm k}s} t}
    J_{m}(qr)
    b_{\mathbf{k}}^s.
    \end{aligned}
	\label{out:cylinder_basis}
\end{equation}

Comparison between cylindrical~(\ref{out:cylinder_basis}) and plane~(\ref{out:helical_press}) wave expansions, leads to the following relation between their amplitudes: 
\begin{equation}
	b_{\boldsymbol \chi}
    =
    \frac{i^{m+1}s}{2^{5/2}\pi} \intop_{-\pi}^\pi \frac{d\phi}{2\pi}e^{-im\phi}a^{s}_{\bm k},
    \quad 
    {\bm k } = \left(q\cos\phi,q\sin\phi,k_z\right).
\label{out:amplitudes}
\end{equation}
In order to prove (\ref{out:amplitudes}), one should pass to integration in~(\ref{out:helical_press}) over cylindrical components of vector ${\bm k}$, then express $a_{\bm k}^s$ in the Fourier series over the angle $\phi$ in plane $(k_x,k_y)=q(\cos\phi,\sin\phi)$ in order to restore $J_m(qr)$ via the integral over $\phi$.

The relation a(\ref{out:amplitudes}) allows one to express the correlation function from Eq.~(\ref{out:ensemble}) in terms of $b_{\bm \chi}$:
\begin{gather}\nonumber 
	\left\langle b_{\boldsymbol{\chi}}\,b_{\boldsymbol{\chi}^{\prime}}^{\ast}\right\rangle =
 2\pi\delta^{ss^{\prime}}\delta_{mm^{\prime}}\delta\left(k_{z}-k_{z}^{\prime}\right)\frac{\delta\left(q-q^{\prime}\right)}{q}  \big\langle |b_{\boldsymbol \chi}|^2\big\rangle ,
    \\ \label{b-statistics}
    \big\langle |b_{\boldsymbol \chi}|^2\big\rangle 
    = 
    \frac{3f}{2^4\Omega}k\Xi(k),
\end{gather}
where, we recall, $k^2 = q^2 + k_z^2$. The wave field is formed only by those waves, whose turning point $r_t = |m|/q$ is closer to the axis $Oz$ than the current coordinate $r_t < r$. Other harmonics are reflected at larger distances, so they no longer have spatial oscillating behaviour in the vicinity of the point $r$ but are decaying toward the axis.

A standing cylindrical wave is the superposition $J_m=(H_m^{(1)}+H_m^{(-1)})/2$ of two traveling waves of equal amplitude, propagating in opposite directions along the radial coordinate, where $H_m^{(\pm1)}$ are the Hankel functions of the first and second kind, respectively. A converging cylindrical wave corresponds to the solution $H_m^{(\sigma)}$, where $\sigma$ is the sign of the frequency $\omega$. The radial component of the energy flux density produced by a separate converging cylindrical wave at a distance $r>r_t$ is equal to
\begin{equation}
	j^r_{\boldsymbol \chi}
    = 
    \frac{1}{2}\mathrm{Re}\, p v^{\ast} 
    =
    -\frac{2\Omega|k_z|}{\pi k^3 r} \big|b_{\boldsymbol \chi}\big|^{2}<0.
\label{out:en_flux}
\end{equation} 
As it should be in the inviscid limit, the product $r j^r_{\boldsymbol \chi} $  does not depend on $r$, see e.g. Ref.~\cite{ivchenko2025absorption}. The total energy flux $2\pi r j^r$ is directed inwards and is produced by the ensemble of those converging cylindrical waves at $R_0$, whose turning point is closer to the origin, $r_t<R_0$:
\begin{equation}
	-2\pi r
     \sum\limits_{s=\pm1}
     \int  \frac{dk_{z}}{2\pi}\intop_{0}^{\infty} qdq\sum\limits_{m=-m_{\scriptscriptstyle{0}}}^{m_{\scriptscriptstyle{0}}}
    \big\langle j^r_{\boldsymbol \chi} \big\rangle 
    =
    2\pi R_0 f,
\label{out:j-total}
\end{equation}
where $m_0 = q R_0 $. To obtain the right-hand side of Eq.~(\ref{out:j-total}), one should first substitute in~(\ref{out:en_flux}) the expression~(\ref{b-statistics}) for $\langle |b_{\boldsymbol \chi}|^{2}\rangle$, implement summation over $m$, then pass from $q=k\sin\theta$, $k_z =k \cos \theta $ to spherical coordinates $\left(k,\theta\right)$ and use the normalization condition in Eq.~(\ref{out:ensemble}). Thus, the definition of the ensemble statistics Eq.~(\ref{out:ensemble}) is indeed consistent with Eq.~(\ref{gen:epsilon}).

In the inviscid limit, a converging cylindrical wave carries a constant incident flux of angular momentum~(\ref{cyl:const_flux}),
\begin{equation}
	r^2\Pi_{{\boldsymbol \chi}}^{r\varphi}
    =
    \frac{r^{2}}{2}\mathrm{Re}\,uv^{\ast}
    =
    \frac{-m\sigma}{\pi k^{2}}\left|b_{\boldsymbol \chi}\right|^{2}.
\label{out:extr_flux}
\end{equation}
This is true up to the radius of absorption (at the critical layer) or reflection (at the turning point), $r>r_\ast,r_t$.

\section{Reynolds shear stress}
\label{sec:uv}
Next, we consider the expression for the Reynolds shear stress, that appears in~(\ref{gen:mean_flow}) in the form of the derivative of angular momentum flux~(\ref{cyl:const_flux}). The waves that were reflected before absorption do not contribute to this average, since the amplitude of the reflected wave are equal to the amplitude of the incident one in the inviscid limit. Only those waves contribute to $\langle u^r u^\varphi\rangle$ that have been absorbed in the critical layer provided the absorption has not yet occurred, $r>r_\ast$. Hence
\begin{equation}
    \begin{aligned}
	 \Pi^{r\varphi}
     =
     \int&\frac{dk_{z}}{2\pi}\intop_{0}^{\infty}qdq 
     \sum_{s,m}\!\!\!\!\!\!{\phantom{\Pi}}^\prime\ 
     \big\langle \Pi_{{\boldsymbol\chi}}^{r\varphi}\big\rangle\,
     \\
     & 
     \theta\left(\left|\omega_{\boldsymbol\chi}\right|-\sigma mW\right)
     \theta(r_\ast - r_t)
     \end{aligned}
    \label{uv:flux_diff}
\end{equation}
The prime symbol at the summation sign means the sum is taken only over those $s,m$ for which $\sigma m W>0$ (i.e., $m s k_z W>0$). This condition means that the converging wave should carry angular momentum of the same sign as that of the vortex, see~(\ref{out:extr_flux}). The first $\theta$-function enforces $r_\ast<r$ (recall that $W(r)$ is presumed to be a monotonic function). The second $\theta$-function retains only those waves whose turning point lies closer to the axis that the critical layer, $r_t<r_\ast$.  Due to the presence of the $\theta$-functions, the order of integration and summation in~(\ref{uv:flux_diff}) can be chosen arbitrarily. We pass again to spherical coordinates in integration, introducing $\zeta = \cos\theta$. Taking into accout~Eqs.~(\ref{b-statistics},\ref{out:extr_flux}), the derivative of the angular momentum flux can be written in the form
\begin{equation}\label{uv:derivative}
    \begin{aligned}
        (r^2 \Pi^{r\varphi})^\prime 
        =
        \frac{3fW^{\prime}(r)}{32\pi^{2}\Omega}
        \int\limits_0^\infty k^2 dk\frac{\Xi(k)}{k} \sum_{s,m} m^2 
        \int\limits_{-1}^1 d\zeta  
        \\
        \times\delta\left( 2\Omega s \zeta-mW\right)
        \theta(r-r_{t}).
    \end{aligned}
\end{equation}
Here $\delta$-function sets the resonance condition $2\Omega s\zeta =mW(r)$, according to which only those waves make a non-zero contribution to~(\ref{uv:derivative}), which have a critical layer at the current position $r=r_\ast$. The requirement~$\theta (r-r_t)=1$ is equivalent to the condition that $k_r^2(\xi)>0$ in~(\ref{cyl:WKB-disp}) for all $\xi>r$. The sum has a cutoff at maximum possible $|m|=m_\ast$, where this condition still holds. The cutoff $m_\ast$ depends on only $k$ due to $\zeta$ is proportional to $m$. We approximate the sum over $m$ with an integral up to $m_\ast$ and obtain
\begin{equation}
	(r^2 \Pi^{r\varphi})^\prime 
    =
    \frac{fW^{\prime}(r)}{16\pi\Omega^2}\intop_{0}^{\infty}\frac{k^{2}dk}{2\pi^{2}}\,
    \frac{\Xi(k)}{k}\, m_{\ast}^{3}.
\label{uv:derivative_sum}
\end{equation}
An explicit expression for $m_\ast$ follows from the requirement that the incident wave is not reflected, $k_r^2(\xi)>0$. It is equivalent to the inequality
\begin{equation}
    \zeta^{2}<
    \zeta_{max}^2,
    \quad 
    \zeta_{max}^2
    =
    \frac{2\widetilde{\Omega}\big(2\widetilde{\Omega}+\Sigma\big)(\xi)\cdot W^2}{\left(2\Omega\right)^{2}\left[W-W(\xi)\right]^{2}}
    \cdot 
    \frac{\left(k\xi W/2\Omega\right)^{2}}{1+\left(k\xi W/2\Omega\right)^{2}}.
\label{uv:zeta_cond}
\end{equation}
The expression in the right hand side of the inequality (\ref{uv:zeta_cond}) can be connected with the cutoff $m_0$ in (\ref{out:j-total}) in the following way. We set $\xi=R_0$, where $W(\xi)/2\Omega = 0$, so we keep all the free waves which are not reflected at distance $R_0$ or further from the axis. We choose any $r<R_0$ such that $W(r)\neq 0$ but $k\xi W(r)/2\Omega\ll1$. Then the first multiplier in (\ref{uv:zeta_cond}) is equal to one, and the denominator in the second multiplier is one as well. Finally we use the resonance equation and arrive to the condition~$\left(kR_0\sin\theta\right)^2\geqslant m^2$, which was used to determine the cutoff $m_0$ during the evaluation in Eq.~(\ref{out:j-total}). Now assume that $r$ is anywhere inside the vortex body. The cutoff value~$ \zeta_\ast^2$ is determined as the smallest between $1$ and the minimum of $\zeta_{max}^2(\xi)$ in the inequality~(\ref{uv:zeta_cond}) which is reached at some in $\xi_\ast$. The radius $\xi_\ast(r)$ has the meaning of a turning point for the harmonic in the ensemble that follows the limiting $m_\ast$, so one has $k_r^2(\xi_\ast)=0$ at $m^2=m_\ast^2$. Approximating $2\widetilde{\Omega}(2\widetilde{\Omega}+\Sigma)(\xi_\ast)\approx (2\Omega)^2$ due to the Rossby number for the vortex is small at the distances, one finds the equation for the extremum's position $\xi_\ast(r)$ in the form
\begin{equation}
	\left[1+\left(k\xi_\ast W/2\Omega\right)^{2}\right]\xi_\ast W^\prime(\xi_\ast)=W(\xi_\ast)-W(r).
    \label{uv:extremum})
\end{equation}
The cutoff $m_\ast$ should be obtained via the resonance relation $\zeta_\ast^2=\left(m_\ast W/2\Omega\right)^2$.

For a monotonically decaying profile $W(r)$ of the angular velocity in vortex, the solution of~(\ref{uv:extremum}) for the extremum definitely lies outside: $\xi_\ast>r$. Result~(\ref{uv:derivative_sum}) is not applicable near the vortex axis, where the viscous absorption mechanism loses its efficiency because of the proximity of a critical layer to a turning point. Comparing the scale of the distance $r-r_t$ with the characteristic length of wave absorption (see the discussion at the end of Section~\ref{sec:cyl}) for harmonics of order $m_\ast$ that dominate in the sum in Eq.~(\ref{uv:derivative}) requires consideration at a sufficient distance from the center:
\[
	r\gg r_\nu=\left(\frac{\nu k_f^2}{W}\right)^{1/3}\frac{m_\ast^{2/3}}{k_f}.
\]

\section{Coherent cyclone profile}
\label{sec:mean}

\begin{figure*}[t]
\centering   

\begin{picture}(\textwidth,300)

\put(0.00\textwidth,200){\includegraphics[width=0.3\linewidth]{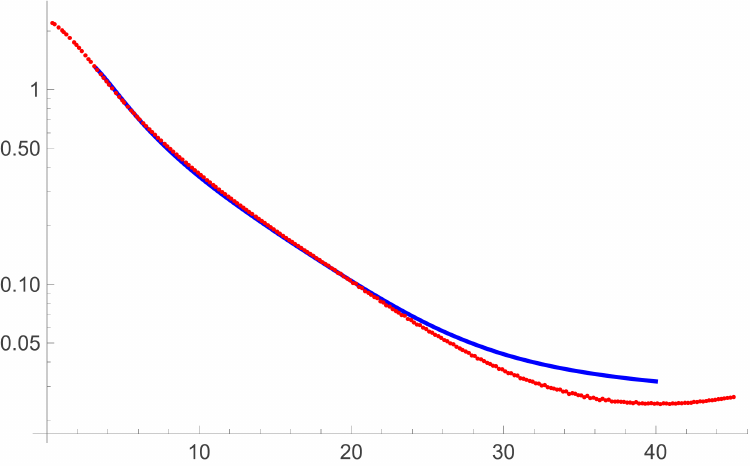}} 

\put(0.35\textwidth,200){\includegraphics[width=0.3\linewidth]{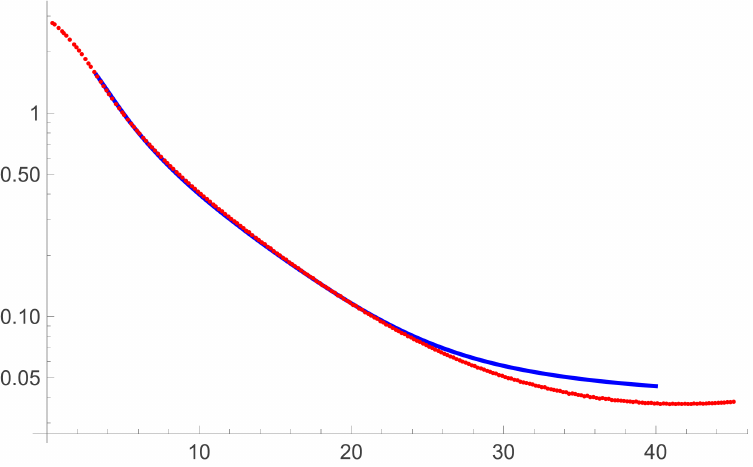}}

\put(0.70\textwidth,200){\includegraphics[width=0.3\linewidth]{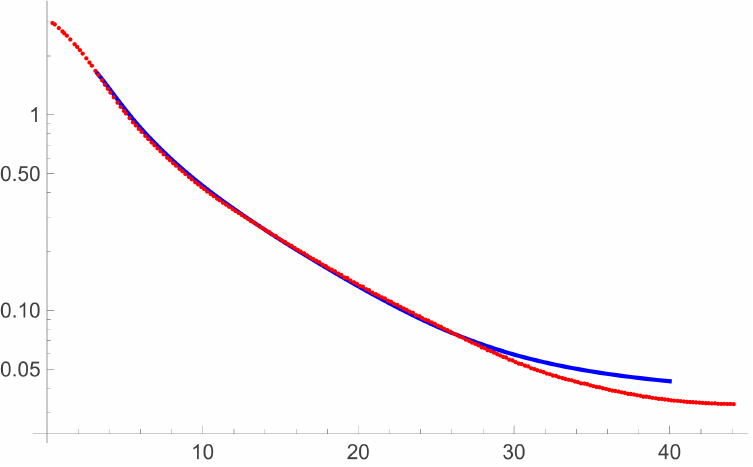}}

\put(0.00\textwidth,100){\includegraphics[width=0.3\linewidth]{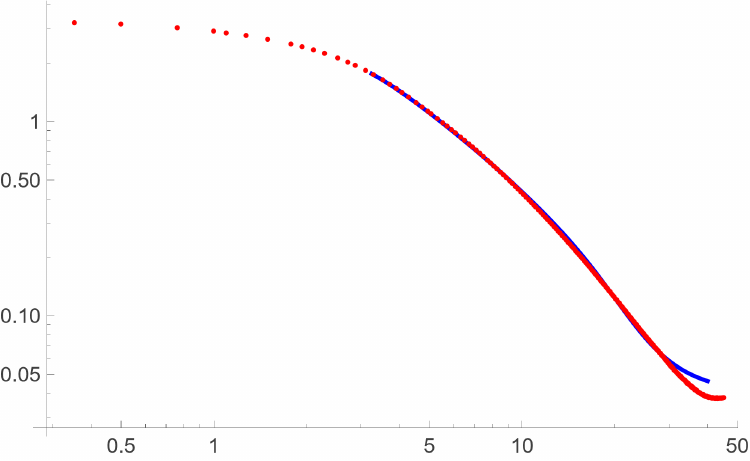}}

\put(0.35\textwidth,100){\includegraphics[width=0.3\linewidth]{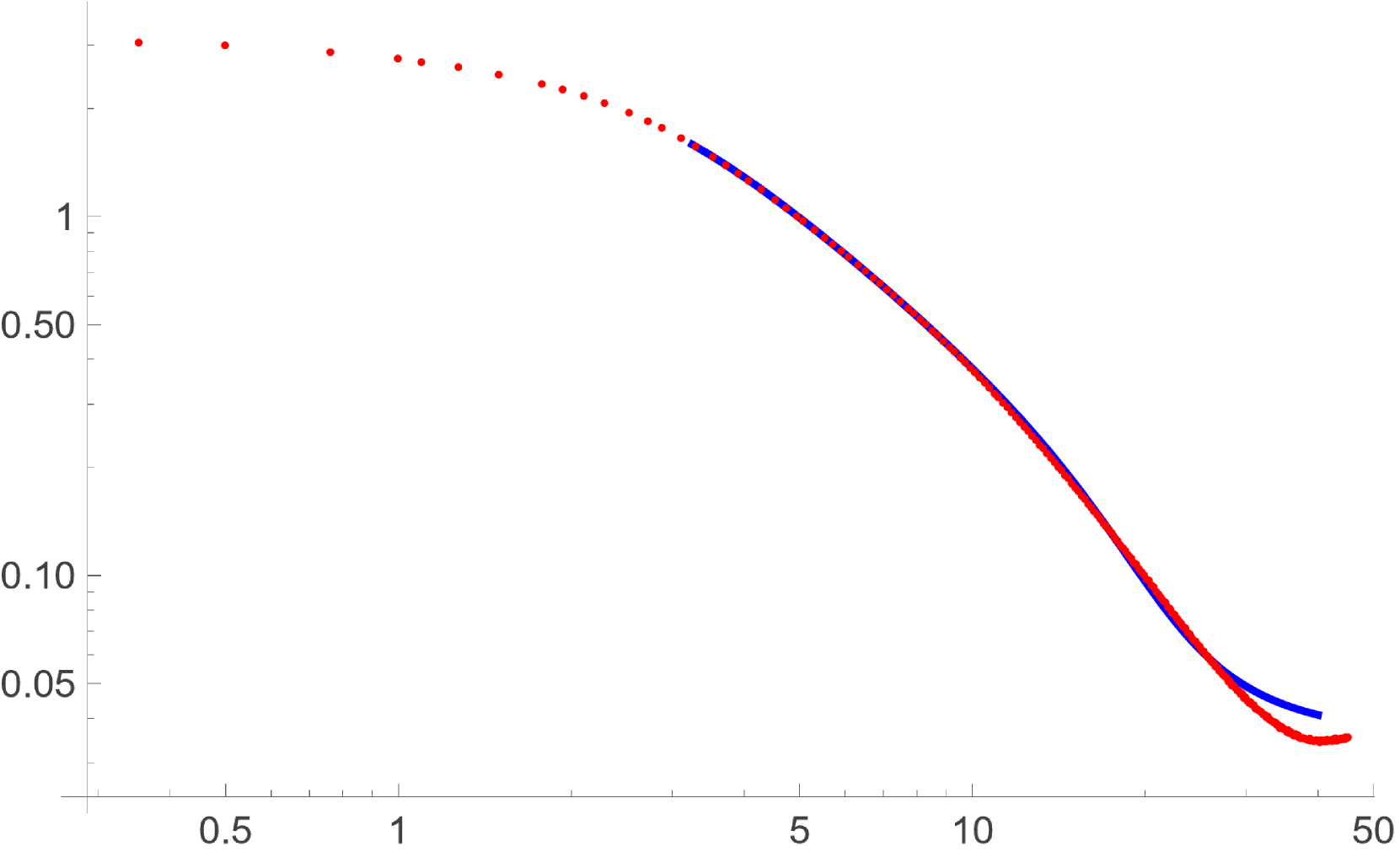}}

\put(0.70\textwidth,100){\includegraphics[width=0.3\linewidth]{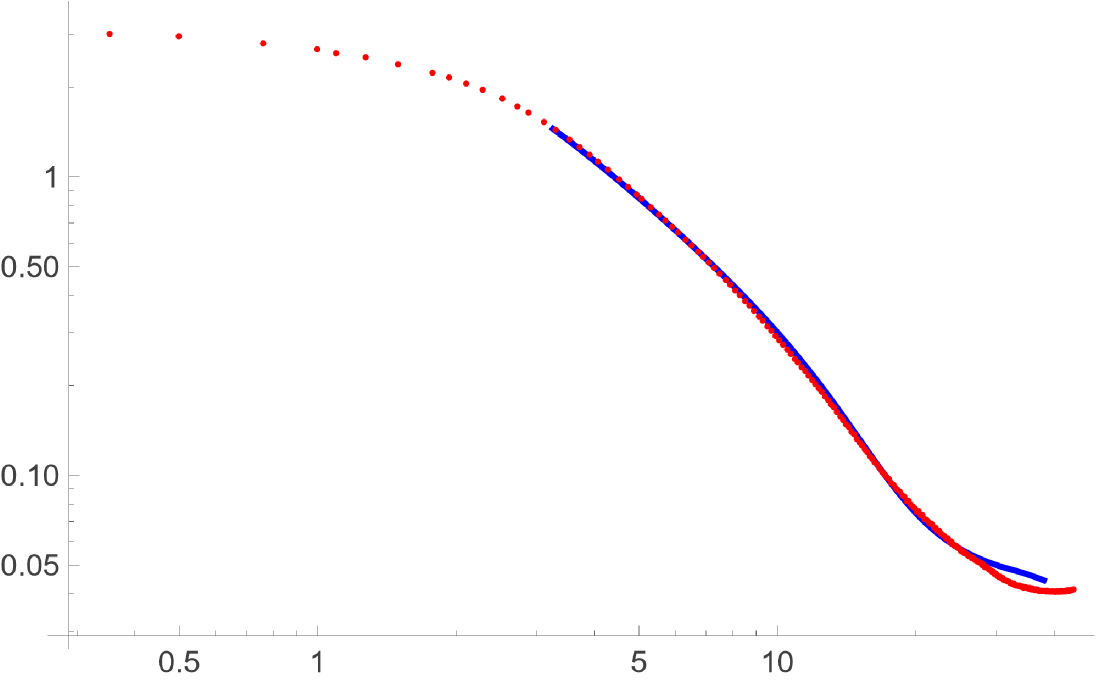}}

\put(0.00\textwidth,  0){\includegraphics[width=0.3\linewidth]{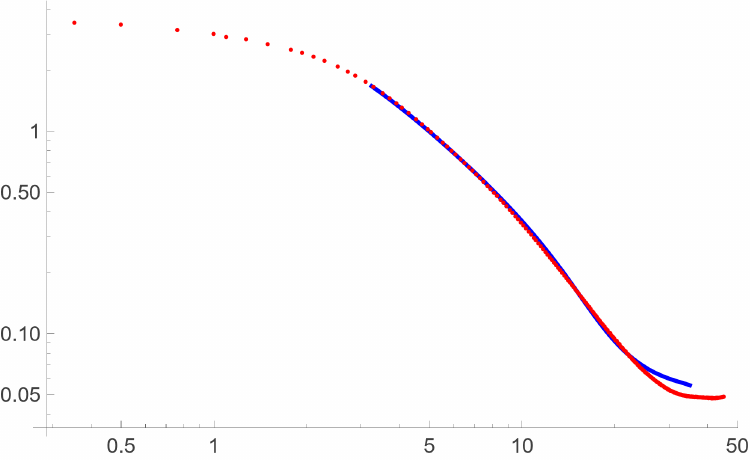}}

\put(0.50\textwidth, 50){
\begin{minipage}[t]{0.5\textwidth}
\begin{tabular}{|c|c|c|c|c|c|c|c|}
\hline  \ensuremath{\Omega}, \ensuremath{\,\mathrm{s}^{-1}}  & 	1.13  &  1.89  &  2.26  &  2.63  &  3.20  &  3.77 & 4.52
 \\
 $R_\mathrm{v}$, cm	 & 	41  &  42  &  43  & 43 & 41 & 39 & 38
 \\
 $\kappa $, \ensuremath{\text{cm}^{-4}}  & 	0.03  &  0.10 &  0.14  &  0.21  &  0.53  & 1.43 & 1.52
 \\
 $k_f$, \ensuremath{\text{cm}^{-1}} & 0.28 & 0.35 & 0.44 & 0.44 & 0.55 & 0.61 & 0.67 
 \\
$\varkappa$,    \ensuremath{\text{cm}^{-1}} & 0.09 & 0.09 & 0.08 & 0.10 & 0.11  &  0.14  & 0.13
 \\
 \hline
\end{tabular}
\refstepcounter{table}\label{table:01}
\vskip10pt 
Table\,\thetable \ Fitting parameters for the numerical simulations. Bottom line: decrement $\varkappa$ values in exponential fit of~$\Sigma(r)$.
\end{minipage}
}

\end{picture}

\caption{Radial dependence~$W(r)$ for cyclones observed in the experiment (dotted lines) and comparisons with numerical solutions of Eq.~(\ref{mean:vort-body}) (solid lines).
}
\label{fig:cycl}

\end{figure*}

Now, we consider the solution of Reynolds equation~(\ref{gen:mean_flow}), which angular velocity $W(r)$ is a monotonically decaying function of a constant sign. We identify such profile with the coherent cyclonic flows observed under the experimental conditions of Ref.~\cite{tumachev2023two} and therefore take $W>0$. First, let's  describe the data obtained in the experiment, which will be used to test the theory.  The measurements of the cyclone’s azimuthal velocity were averaged over the angle $\varphi$ and over the lifetime of the vortices. The graphs of $W(r)$ plotted in Fig.~\ref{fig:cycl} are flow profiles averaged over 5 similar long‑lived cyclones observed in a single experimental run. The standard deviation for the profiles is within 10\%. At large distances $W(r)$ exhibits a distinct minimum. The subsequent growth of $W(r)$ occurs at distances on the order of the installation radius (0.5 m). Therefore, we attribute it to the outer mean flow and adopt the position of the minimum $R_\mathrm{v}$ as the boundary of the cyclone. 

For a monotonically decreasing function $W$ it is convenient to use a WKB-type substitution,
\begin{equation}
	W(r)=W(0) e^{-\intop_0^r dx \varkappa(x)},
\label{mean:WKB-cyclone}
\end{equation}
where $\varkappa =-W'(r)/W(r)$ is the local damping decrement. As the flow varies on the scale of the flow, $\varkappa/k_f \ll 1$. The equation~(\ref{uv:extremum}) for the position of the extremum takes the form:
\begin{gather}
	e^A=1+\varkappa(\xi_\ast)\xi_\ast\left[1+B\right],
\nonumber	
	\\
	 A= \intop_r^{\xi_\ast} dx \varkappa(x),
  \quad 
  B=\left(k \xi_\ast W(r)/2\Omega\right)^2.
\label{mean:B-eq}
\end{gather}

The combination $B=\left(k \xi_\ast W(r)/2\Omega\right)^2$ appearing in the condition~(\ref{uv:zeta_cond}) governs the behaviour of the solution of~(\ref{uv:extremum}). It decreases away from the vortex axis. Near the vortex's center, $\xi_\ast(r)$ remains of the order of the vortex size and $W(0)$ is comparable to $\Omega$, so $B\gg 1$. The detailed analysis for dependence $\xi_\ast(r)$ is presented in Appendix~\ref{app:extreme}. It shows, see Eq.~(\ref{app2:xi-prime}), that the minimum moves towards the axis, $d\xi_\ast/dr<0$ if $B>1$, otherwise $\xi_\ast(r)$ grows. For $B\ll 1$ this growth can be approximately considered to be linear. The growth slows down significantly as $\xi_\ast$ approaches the vortex's boundary $R_{\mathrm{v}}$.

The minimal position $\xi_{\ast,min} \approx 2k_f^{-1}\Omega/W(r_{min})$ is achieved at the distance $r_{min}$ where $B\approx 1$. The angular velocity of the cyclone at $r_{min} \sim 10 \,\mathrm{cm}$ is already much weaker than its on-axis value since $\varkappa \,r_{min}\gtrsim 1$, $W(r_{min})\ll W(0)$ (see representative example in Fig.~\ref{fig:extremum} in Appendix~\ref{app:extreme}). Hence, the minimal $\xi_{\ast,min}$ is, in turn, far from the point $r_{min}$ in the sense that $W(\xi_{\ast,min})\ll W(r_{min})$. The function in expression~(\ref{uv:zeta_cond}) $m_{max}^2(\xi)$ only slightly decreases away from the vortex axis when $\xi$ is far from $r$, so that $W(\xi)\ll W(r)$; see Eq.~(\ref{app2:logm-prime}) in Appendix. Therefore, in describing the body of the cyclone, one may approximately assume that the cutoff~(\ref{uv:zeta_cond}) is set by the outer scale~$R_\mathrm{v}$, and the deviation from its value in $\xi_\ast <R_{\mathrm{v}}$ can be neglected.

Let us make one more simplification. We will assume the spectrum of inertial waves to be narrow at the periphery of the vortex flow, so that the correlation function~$\Xi(k)$, see Eq.~(\ref{out:ensemble}), is non-zero only near the characteristic number $k_f$. As a result of all our simplifications, Eq.~(\ref{gen:mean_flow}) for the mean flow becomes an ordinary differential equation in this reduced model. Linear bottom friction may be taken to be the dominant dissipation mechanism for the vortex outside its viscous core having size of order $\sqrt{\nu/\alpha}$. According to the experimental data, the bottom friction dominates over viscosity starting from $r_0 \approx 3 \,\mathrm{cm}$, which is roughly double $\sqrt{\nu/\alpha}$. In Fig.~\ref{fig:cycl}, the profiles based on measurements (points, see details at the beginning of the section) are compared with the numerical solution (solid curve) of the differential equation:
\begin{multline}
	\frac{W^{\prime}}{r^3}\left[\mathrm{min}\left(\left(\frac{2\Omega}{W}\right)^{2},\frac{W^{2}}{\left(W-W_{\ast}\right)^{2}}\frac{\left(k_f R_{\mathrm{v}}\right)^2}{1+B_{\mathrm{v}}^{2}}\right)\right]^{3/2}
	\\ 
	    +\kappa W=0,\quad \kappa=\alpha k_f^3 /F,\quad W_{\ast}=W\left(R_{\mathrm{v}}\right),
\label{mean:vort-body}
\end{multline}
where $F=f k_f^2 /16\pi\Omega^2 $, so the coefficient $\kappa $ has dimension $\,\mathrm{cm}^{-4}$ and the notation~$B_\mathrm{v}=\left(k R_{\mathrm{v}} W(r)/2\Omega\right)^2$ has been introduced.

To solve (\ref{mean:vort-body}) one has to specify the boundary conditions and parameters in the equation. We have developed the following algorithm. The initial condition $W(r_0)$ and the vortex boundary $R_{\mathrm{v}}$ are determined based on the experimentally measured profile. The coefficient $\kappa$ was chosen so that the derivative of the solution at the starting point, $W^\prime(r_0)$, coincided with the experimental profile. The scale $k_f$ of the waves in the ensemble (estimated from experiments~\cite{tumachev2023two,tumachev2024observation}: $k_f\sim 1 \,\mathrm{cm}^{-1}$) was adjusted to reproduce the subsequent behaviour of the profile $W(r)$. The values for each profile are given in Table~\ref{table:01}. The parameters $\kappa$ and $k_f$ increase with an increase in the system’s rotation speed $\Omega$. Number of specified digits for their value is dictated by the best fit of solution~(\ref{mean:vort-body}) to the measured $W(r)$; it is finer than the uncertainty which the experimental dependence $W(r)$ itself is measured with. The deviation of the numerical solution after $r=25\,\mathrm{cm}$ increases as $W(r)$ approaches its peripheral value $W_\ast$. At this distance the crude replacement $\xi_\ast \rightarrow R_{\mathrm{v}}$ in Eq.~(\ref{mean:vort-body}) is no longer applicable since the assumption that the constraint~(\ref{uv:zeta_cond}) on $m^2$ varies slowly with $\xi$ holds no longer (see Eq.~(\ref{app2:m2-prime}) in Appendix~\ref{app:extreme} and the discussion below).

The measured profiles in Fig.~\ref{fig:cycl} supplied with the solutions of Eq.~(\ref{mean:vort-body}) are plotted on a log-log scale for $\Omega>2.5 \,\mathrm{s}^{-1}$ since $W(r)$ displayed behavior resembling a power-law decay. In contrast, for slow rotation $\Omega = 1.1-2.3\, \mathrm{s}^{-1}$ one can identify a region in the profile $W(r)$ where the angular velocity decay approximately exponentially, $W(r)$; such cyclones are plotted in log-scale. However, we observed that $\Sigma(r)$ has exponential intermediate asymptotic at all the rotation speeds, $\Sigma(r)\propto e^{-\varkappa r}$, see an example in Fig.~\ref{fig:profile}. The values of $\varkappa$ corresponding to the asymptotics are listed in~Table~\ref{table:01}. In average, $\varkappa$ slightly increases with the rotation speed $\Omega$.

One can calculate the dependence $\epsilon(\Omega)$ with the parameters from the~Table~\ref{table:01}, using~(\ref{gen:epsilon},\ref{mean:vort-body}). They are scattered irregularly over the interval $\left( 0.06,0.11\right) \mathrm{cm}^2/\mathrm{s}^3$, which reflects the actual accuracy in estimation of parameters. From the experiment one can compute directly $\epsilon_{\mathrm{exp}}=2\alpha E $, where $E$ is the mean energy density of the flow. Both values turn out to be close to each other (see Fig.~\ref{fig:epsilon} in Appendix) with mean ratio $\epsilon/\epsilon_{\mathrm{exp}}\approx 1.4$, indicating that the theory developed here is reasonable.

\section{Conclusions and discussion}
\label{sec:conc}

We have developed a theory describing the structure of long-living geostrophic vortices formed in a turbulent flow of rotating fluid. It is based on linear approximation when describing an inertial wave propagation on the background of the vortex. It was assumed that the waves are produced by a statistically isotropic forcing at the periphery region of the vortex. The theory has been successfully applied to describe vortices that were observed in experiment. The results are reasonably consistent with independent measurements of the excitation power supporting the flow.

At lower rotation speeds, the linear approximation for wave dynamics ceases to be valid. A sign of the growing role of nonlinear wave interaction is the increasingly less steep energy spectrum as the rotation speed of the cube decreases. A qualitative change in the regime of inertial wave dynamics is accompanied by the establishment of the large anticyclone locating at the center of the flow. Our additional analysis of the experimental data \cite{tumachev2024observation} shows that the energy spectrum is $E_k\propto k^{-5/2}$ at the periphery of the anticyclone, where the mean flow is cyclonic one. (The spectrum presented in \cite{tumachev2024observation} describes the inner anticyclonic region of the flow.) In addition, the first of the conditions (\ref{gen:inequality2}) is violated in the region. For example, at a rotation rate of $\Omega\approx 0.38\,\text{s}^{-1}$, the shear rate is $\Sigma \approx 0.1\,\text{s}^{-1}$, while the rate of nonlinear wave interaction is $1/t_{tr} = \langle \omega_z^2\rangle/2\Omega \approx 0.3\,\text{s}^{-1}$, where $\langle \omega_z^2\rangle$ is the root-mean-square of the fluctuating part of the vertical vorticity. The hierarchy of rates $2\Omega > 1/t_{tr} > \Sigma$ implies that there is a direct cascade in weak inertial-wave turbulence \cite{galtier2003weak,gelash2017complete} perturbed by an imposed relatively weak large-scale shear flow. The significance of nonlinear wave–wave interaction also follows from the theory developed here. Equation (\ref{mean:vort-body}) cannot be satisfied if $W(r)$ changes its sign. Thus, a theory going beyond the linear description of waves is required to describe the anticyclone.

\section{Acknowledgments}

The experimental part of this work was carried out by all the authors in Landau ITP and was supported by the Russian Science Foundation, Grant No. 23-72-30006. The analytical theory was developed by NAI and SSV, who were supported by the Basic research program of HSE, Project No. HSE-BR-2025-57.

\bibliography{inertial-waves}

\appendix

\section{Detailed analysis of the cutoff dependence $m_\ast(r)$ for the azimuthal number.}
\label{app:extreme}
In this Section we consider the minimum, which  point is given by~(\ref{uv:extremum}). The equation yields the expression how moves the position $\xi_\ast(r)$ (the notation for $B$ was introduced in~(\ref{mean:B-eq})):
\begin{equation}
   \frac{d\xi_{\ast}(r)}{dr}=\frac{-W^{\prime}(r)\left[1+2B \xi_{\ast}W^{\prime}(\xi_{\ast})/W(r)\right]}{\left(1+B\right)\xi_{\ast}W^{\prime\prime}(\xi_{\ast})+3BW^\prime(\xi_{\ast})}.
\label{app2:xi-prime_gen}
\end{equation}

\begin{figure*}[t]
    \centering
      \includegraphics[scale=0.3]{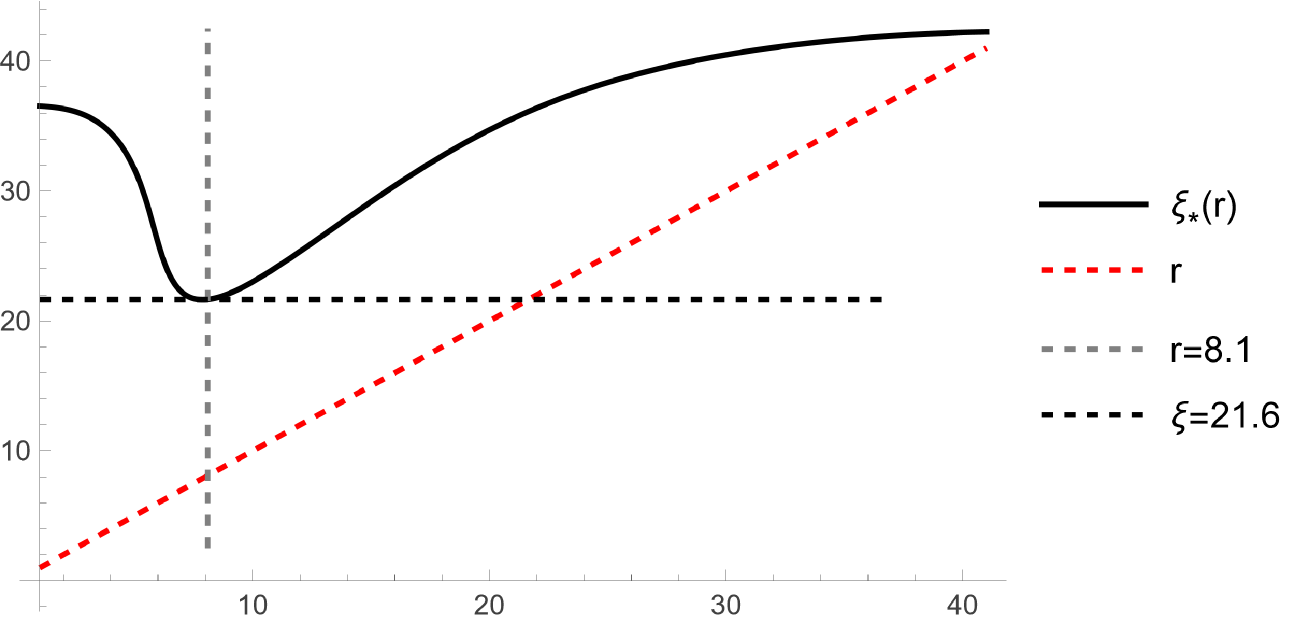}
      \hfill
      \includegraphics[scale=0.35]{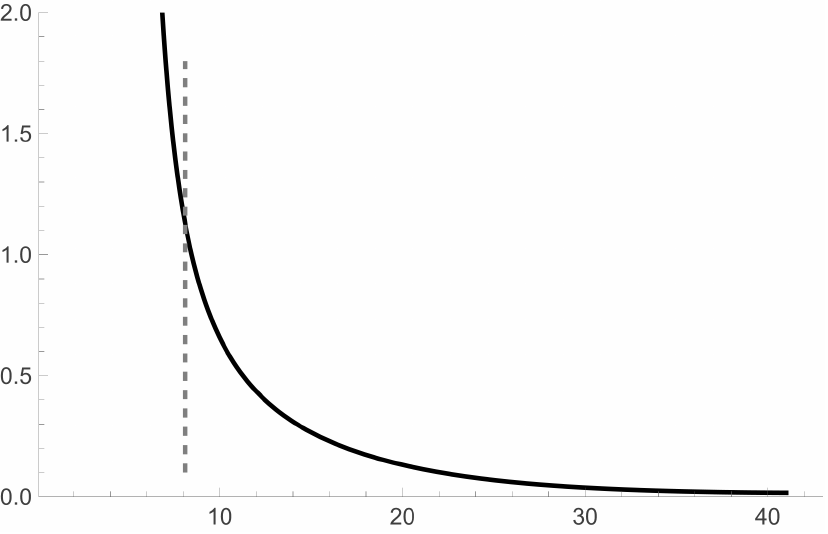}
      \hfill
      \includegraphics[width=0.3\linewidth]{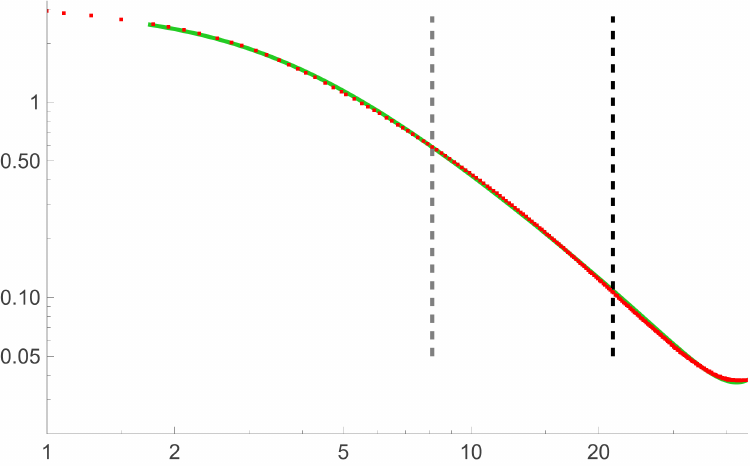}
    \caption{Extremum position $\xi_\ast$ of the expression in~(\ref{uv:zeta_cond}) and the value of $B$, see~(\ref{mean:B-eq}), for cyclone at $\Omega=2.63\,\text{s}^{-1}$. Its $W(r)$ profile (on the right, red dots) was fitted on interval~$(2,44)\,\text{cm}$, after that the fit (green line) was taken for numerical solution of minimum in Eq.~(\ref{uv:zeta_cond}) (the wavenumber $k\rightarrow k_f=0.44\, \text{cm}^{-1}$). Dashed lines on the first plot corresponds to the minimal position $\xi_\ast (r_{min})=21.6$ cm (black), which is achieved at~$r_m=8.1$~cm (gray).}
\label{fig:extremum}
\end{figure*}

\begin{figure*}[t]
   \centering
    \includegraphics[width=0.4\linewidth]{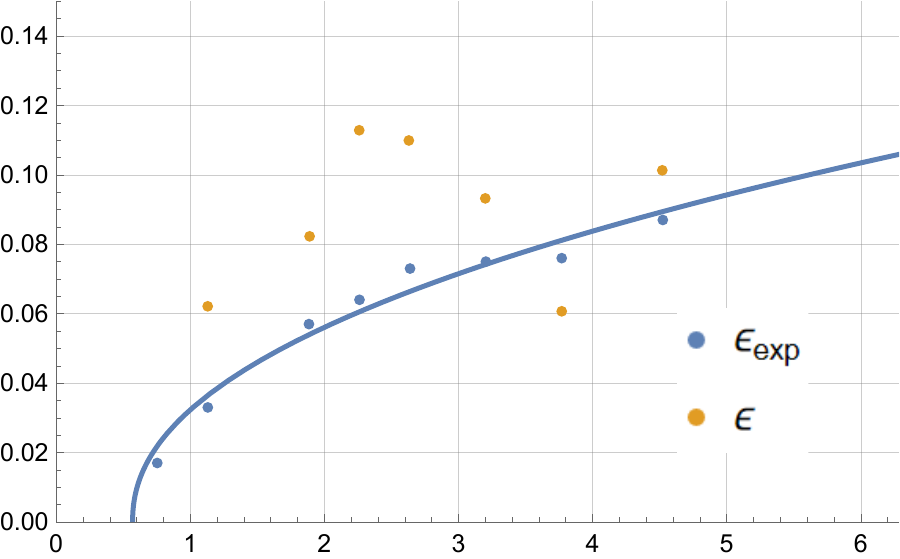}
    \caption{The experimental values of the mean volume power $\epsilon_{exp}$ produced by the mixers and its value $\epsilon$ extracted from the result of numerical solution of Eq.~(\ref{mean:vort-body}). The fit for the experimental data is $\epsilon_{exp} = 0.048\cdot(\Omega - 0.57)^{0.46}$, all the physical quantites are measured in CGS-units.}
    \label{fig:epsilon}
\end{figure*}

Assuming $\xi_\ast$ is at a distance, that following conditions hold:
\[
W(\xi_{\ast})\ll W(r),\quad\varkappa(\xi_\ast)\xi_\ast \gg 1,
\]
one derives an approximate formula
\begin{equation}
    \frac{d\xi_{\ast}(r)}{dr}\approx\frac{\varkappa(r)}{\varkappa(\xi_\ast)}\cdot\frac{1-B}{1+B},
\label{app2:xi-prime}
\end{equation}
where $\varkappa(r)$ is a local damping decrement which was introduced in~(\ref{mean:WKB-cyclone}). Farther, when $\xi_\ast$ is near the minimum of angular velocity $W$ at $R_{\mathrm{v}}$, the asymptotics of~(\ref{app2:xi-prime}) changes to a slower motion: from the general formula~(\ref{app2:xi-prime_gen}) it follows that
\[
\frac{d\xi_{\ast}(r)}{dr}\approx-\frac{W^{\prime}(r)}{R_{\mathrm{v}}W^{\prime\prime}\left(R_{\mathrm{v}}\right)}
\]
i.e. $\xi_\ast^\prime $ is small in the parameter $\varkappa(r)/R_{\mathrm{v}} \varkappa^\prime(R_{\mathrm{v}})\ll 1$ (because $R_{\mathrm{v}}$ is the largest scale, the edge of the vortex).

Along with that the constraint of $m^2$ given by~(\ref{uv:zeta_cond}), varies slowly with~$\xi$ away from the vortex's center. The expression for logarithmic derivative~($2\widetilde{\Omega}(2\widetilde{\Omega}+\Sigma)(\xi_\ast)\approx (2\Omega)^2$ there) is 
\begin{equation}
    \frac{\partial \ln m^{2}}{\partial\xi}\approx 2\left[\frac{W^{\prime}(\xi)}{\left(W-W(\xi)\right)}+\frac{1}{\xi}\frac{1}{1+B}\right]
\label{app2:m2-prime}
\end{equation}
and implies the characteristic scale of its change as
\[
\mathrm{min}\left[(1+B)\xi,\varkappa(\xi)^{-1} (W-W(\xi))/W(\xi)\right].
\]
Its minimum value at $\xi=\xi_\ast(r)$, which defines the cutoff for sum over harmonics, satisfies the equation:
\begin{equation}
    \frac{d\ln m_\ast}{dr}=\frac{W^\prime}{W}\cdot\frac{\xi_\ast W^\prime(\xi_\ast)-W(\xi_\ast)}{W-W(\xi_\ast)}.
\label{app2:logm-prime}
\end{equation}

\end{document}